# DeepRT: deep learning for peptide retention time prediction in proteomics

Chunwei Ma[1,2#], Zhiyong Zhu[3#], Jun Ye[3], Jiarui Yang[1,4], Jianguo Pei[3], Shaohang Xu[1,2], Ruo Zhou[1,2], Chang Yu[3], Fan Mo[3], Bo Wen[1,2*], Siqi Liu[1,2*]

1 BGI-Shenzhen, Shenzhen 518083, China.
2 China National GeneBank, BGI-Shenzhen, Shenzhen 518083, China.
3 Intel Asia-Pacific Research & Development Ltd, Shanghai 200241, China.
4 Qingdao University-BGI Joint Innovation College, Qingdao University, Qingdao 266071, China.
# Equal contribution
* Correspondence:
Dr. Siqi Liu, BGI-Shenzhen, Shenzhen, 518083, China, E-Mail: siqiliu@genomics.cn
Bo Wen, BGI-Shenzhen, Shenzhen, 518083, China, E-Mail: wenbo@genomics.cn

**ABSTRACT**
**Summary:** Accurate predictions of peptide retention times (RT) in liquid chromatography have many applications in mass spectrometry-based proteomics. Herein, we present DeepRT, a deep learning based software for peptide retention time prediction. DeepRT automatically learns features directly from the peptide sequences using the deep convolutional Neural Network (CNN) and Recurrent Neural Network (RNN) model, which eliminates the need to use hand-crafted features or rules. After the feature learning, principal component analysis (PCA) was used for dimensionality reduction, then three conventional machine learning methods were utilized to perform modeling. Two published datasets were used to evaluate the performance of DeepRT and we demonstrate that DeepRT greatly outperforms previous state-of-the-art approaches ELUDE and GPTime.
**Availability:** DeepRT software and its user manual are freely available at: https://github.com/horsepurve/DeepRT and are provided under a GPL-2 license.
**Contact:** siqiliu@genomics.cn
**Supplementary information:** Supplementary data are available online.

## 1 INTRODUCTION

In liquid chromatography-tandem mass spectrometry (LC-MS/MS) based proteomics experiments, peptides are usually separated by liquid chromatography before being introduced into mass spectrometry. The time that records the moment for a peptide eluted from LC and injected into mass spectrometer is referred to retention time (RT). As RT is relied on the physical and chemical properties of a peptide on a certain LC system, it is predictable in theory and reproducible in experiment (Moruz and Käll, 2016). RT is not only an experimental data, but would be also a parameter to assist identifying and targeting peptides, and analyzing with data-independent acquisition (DIA) in proteomics (Moruz and Käll, 2016).

As the peptide RT is useful to guide the annotation of experimental data, how to gain its prediction within a small error range has attracted an attention in algorithm development, such as SSRCalc (Krokhin, 2006), ELUDE (Moruz, et al., 2010) and GPTime (Maboudi Afkham, et al., 2016). SSRCalc takes several parameters into account of RT, like amino acid composition, residue position, peptide length, hydrophobicity, pI, nearest-neighbor effect of charged side chains (K, R, H), and propensity to form helical structures. ELUDE utilizes the machine learning approach, support vector regression (SVR), to estimate RT based on 60 hand-crafted features derived from amino acid composition of a peptide. With the same hand-crafted features used in ELUDE, GPTime applies Gaussian Process for RT prediction. All of the RT predictors reported so far are still not practicable in LC or MS data treatment due to relatively larger errors in estimation. It is generally recognized that the selected peptide features restrict such algorithm development. Since all the hand-crafted features or rules are based upon personal expertise or knowledge, a question is naturally raised whether the peptide features serving for RT prediction could be generated from a machine analysis, a non-personal experience dependent approach.

Hence, we propose DeepRT, a machine learning based software, which utilizes the advanced deep learning framework to extract the peptide features with integration of peptide sequences and their LC behaviors. To our knowledge, DeepRT is the first attempt to apply deep learning method in theoretical estimation of peptide RTs.

## 2 METHODS AND IMPLEMENTATION

Overview of the data processing with DeepRT is illustrated in Figure 1.

### 2.1 Peptide feature extraction through deep learning

DeepRT uses deep CNN and RNN to extract peptide features. CNN layer consists of locally connected neurons that share weights and enable capturing the spatial features of input data (Lee, et al., 2009). Before being fed into CNN, peptide sequences are first converted to two-dimensional matrix using one-hot encoding and padding. DeepRT uses 4 convolutional layers, each with 20 filters and one-dimensional feature map. After treatment of convolutional layers, a dropout layer is added to prevent over-fitting (Srivastava, 2013). Finally, a fully-connected layer performs linear regression for the flattened features. To have efficient gradient propagation, all the layers in CNN use leaky ReLU as activation function that allows 0.01 gradient when the neurons are not active (Glorot, et al., 2011). RNN (Lipton, et al., 2015) views each peptide sequence as a sentence and transform each word (amino acid) into a dense vector with the length of 20, and extract peptide fea-tures using several Long Short-Term Memory (LSTM) layers (Hochreiter and Schmidhuber, 1997), in which the layers are stacked on the top of input data and each one outputs its internal projection with the length of 128. Similar to CNN, a final regression layer is entailed for supervised feature learning. In RNN model, a dropout layer with 0.2 dropout rate is added to all the two LSTM layers.

To unearth as many features as possible, DeepRT runs with many different parameter settings. For CNN model, the initial learning rates are set as 0.01 and 0.001, with filter sizes from 3 to 9 and dropout ratios at 0.2 and 0.5. For RNN model, 1 or 2 hidden layers are chosen, and the regression methods are set as linear or logistic (detailed parameter settings are listed in Supplementary material). Both CNNs and RNNs use stochastic gradient descent (SGD) with 128 batch size. Based on a training dataset, the extraction of peptide features is run by the 7 CNN models and 4 RNN models in parallel, and the feature candidates go through principal component analysis (PCA) to finalize the features with 95% cumulative proportion of explained variance.

## 2.2 Non-linear regression and ensemble

For RT prediction, DeepRT first trains 3 non-linear classifiers upon the learned features, support vector regression (SVR) with parameter gamma = $e^{-9}$ and C = $e^6$, random forest (RF) with 400 trees, and gradient boosting (GB) with 1000 boosting stages. And RT is predicted based on bagging approach which combines the three regression models (Breiman, 1996).

## 2.3 Software implementation

DeepRT is implemented in Python, utilizing Theano (0.9.0 dev1) (Bastien, et al., 2012) and Keras (1.0.1) (https://github.com/fchollet/keras) for deep learning training and feature extraction, and taking Sklearn (0.17.1) (Pedregosa, et al., 2011) for non-linear regression.

## 3 RESULTS AND DISCUSSION

Two publically available datasets were taken for evaluation of the performance of DeepRT and comparison of DeepRT with two existing software, ELUDE and GPTime. Each dataset was split into three independent subsets for training, validating and testing (8:1:1), respectively. DeepRT was run on the Intel Xeon Processor E5-2699 v4 (2.20GHz, 44 cores) with 128G memory, achieving faster training speed than GPU (data not shown). Three performance metrics were used in evaluation and comparison of the predicted results, such as Pearson correlation coefficient (Pearson), root-mean-square error (RMSE), and the minimal time window containing the deviations between actual and predicted RT for 95% of the peptides ($\Delta t_{95\%}$). As shown in Figure 1B, DeepRT achieved much better Pearson correlation at 0.985 and 0.992 between actual and predicted RTs in the two datasets, respectively, while kept much smaller RMSE and $\Delta t_{95\%}$ as compared with ELUDE and GPTime. Remarkably, the performance improvement for dataset 2 was better than that for dataset 1, suggesting that the larger training set was used, the better prediction was achieved.

To our best knowledge, DeepRT is the first approach of RT prediction based on deep learning, which took the proteomics data as the experimental evidence, utilized the state-of-the-art techniques for feature learning and demonstrated the significant improvement in RT prediction. We expect it would facilitate peptide identification in both data-dependent acquisition (DDA) and DIA proteomics studies.

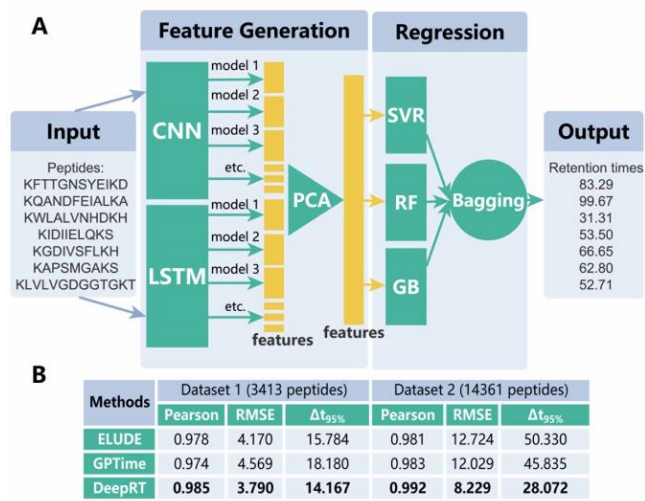

Figure 1. (A) The framework of DeepRT. (B) Comparison of deepRT with ELUDE and GPTime.

*Conflict of Interest*: none declared.